\documentclass[floatfix,a4paper,prl,twocolumn,superscriptaddress,preprintnumbers,amsmath,amssymb]{revtex4}

\usepackage{hyperref}
\usepackage{amsmath}
\usepackage{amssymb}
\usepackage{graphicx}
\usepackage{color}
\usepackage{multirow,ulem}
\usepackage{verbatim}

\renewcommand\approx{%
  {\sim}
}

\begin{document}
\preprint{}

\title{Signatures of single quantum dots in graphene nanoribbons within the quantum Hall regime}

\author{E.~T\'{o}v\'{a}ri}
\affiliation{Department of Physics, Budapest University of Technology and Economics, and Condensed Matter Research Group of the Hungarian
Academy of Sciences, Budafoki \'{u}t 8, 1111 Budapest, Hungary}

\author{P. Makk}
\affiliation{Department of Physics, University of Basel, Klingelbergstrasse 82, CH-4056 Basel, Switzerland}

\author{P.~Rickhaus}
\affiliation{Department of Physics, University of Basel, Klingelbergstrasse 82, CH-4056 Basel, Switzerland}

\author{C.~Sch\"{o}nenberger}
\affiliation{Department of Physics, University of Basel, Klingelbergstrasse 82, CH-4056 Basel, Switzerland}

\author{S.~Csonka}
\email{csonka@mono.eik.bme.hu}
\affiliation{Department of Physics, Budapest University of Technology and Economics, and Condensed Matter Research Group of the Hungarian
Academy of Sciences, Budafoki \'{u}t 8, 1111 Budapest, Hungary}

\date{\today}

\begin{abstract}
We report on the observation of periodic conductance oscillations near quantum Hall plateaus in suspended graphene nanoribbons. They are attributed to single quantum dots that form in the narrowest part of the ribbon, in the valleys and hills of a disorder potential. In a wide flake with two gates, a double-dot system's signature has been observed. Electrostatic confinement is enabled in single-layer graphene due to the gaps that form between Landau levels, suggesting a way to create gate-defined quantum dots that can be accessed with quantum Hall edge states.
\end{abstract}


\maketitle

{\it Introduction---}
The Dirac spectrum results in several peculiar features in the charge transport of graphene, such as Klein tunneling, or the special Berry phase and the half-integer quantum Hall-effect \cite{Katsnelson2006, Novoselov2005, Zhang2005, Neto2009}. The high mobility of graphene offers a good platform for field effect transistors, whereas the low spin-orbit coupling \cite{Gmitra2009} and small amount of $^{13}\rm C$ nuclear spins make it promising for the realization of long-lifetime spin qubits \cite{DiVincenzo1998, Loss1998, Trauzettel2007, Droth2015}. However, from an application point of view, the absence of a band gap places limitations: it hinders effective electrostatic confinement of electrons, which makes the fabrication of spin qubits challenging and results in high OFF state currents for field effect transistors. 

Creating nanoribbons in graphene provides a way to generate a band gap due to one dimensional confinement \cite{Son2006, Son2006a}. The common technique to confine electrons in a graphene quantum dot (QD) or ribbon is based on tailoring the graphene sheet by etching. In QD devices \cite{Stampfer2008, Liu2009, Moriyama2009, Schnez2010} thin graphene nanoribbon sections play the role of tunnel barriers. Promising results have been achieved, e.g. detection of the QD's orbital spectrum \cite{Ponomarenko2008, Liu2010}, or observation of the spin-filling sequence \cite{Guettinger2010}. However, edge roughness, inhomogeneities in the substrate, fabrication residues, and the unpredictability of the nanoribbons that act as tunnel barriers place clear limitations to this technology \cite{Han2007, Stampfer2009, Han2010, Gallagher2010}.

Other confinement strategies involve opening a gap in bilayer graphene using perpendicular electric fields, or exploiting the angle-dependent transmission in p-n junctions. Both techniques require ultra-clean high mobility junctions, for which encapsulation in hBN \cite{Dean2010,Mayorov2011} or suspension of the graphene flake \cite{Bolotin2008, Tombros2011} is required. Recently quantum dots and point contacts have been created by utilizing the gap opening in bilayer graphene \cite{Allen2012, Goossens2012, Varlet2014}. Futhermore, beamsplitters and waveguides were fabricated using p-n junctions \cite{Rickhausguide2015, Rickhausbeam2015}, however, the confinement offered by the p-n transition is soft and electrons can leak out.

In this paper we focus on a different method, which uses magnetic fields to form a gap in the bulk of single-layer graphene. Applying a perpendicular magnetic field $B$, Landau levels (LLs) form with remarkably high energy spacing: for example, the energy of the fourfold degenerate, $N=1$ LL is $36.3~ \rm meV \it \cdot \sqrt{B \rm ~[T]}$. Combining this $B$ field-induced gap with a local electrostatic field, a confinement potential for quantum dots can be achieved, which can be read out via edge states. In this work we present the transport characterization of suspended single-layer graphene strips where single and double dots form based on this principle.

\begin{figure} [t!]
\centering
\includegraphics[width=0.5\textwidth,clip]{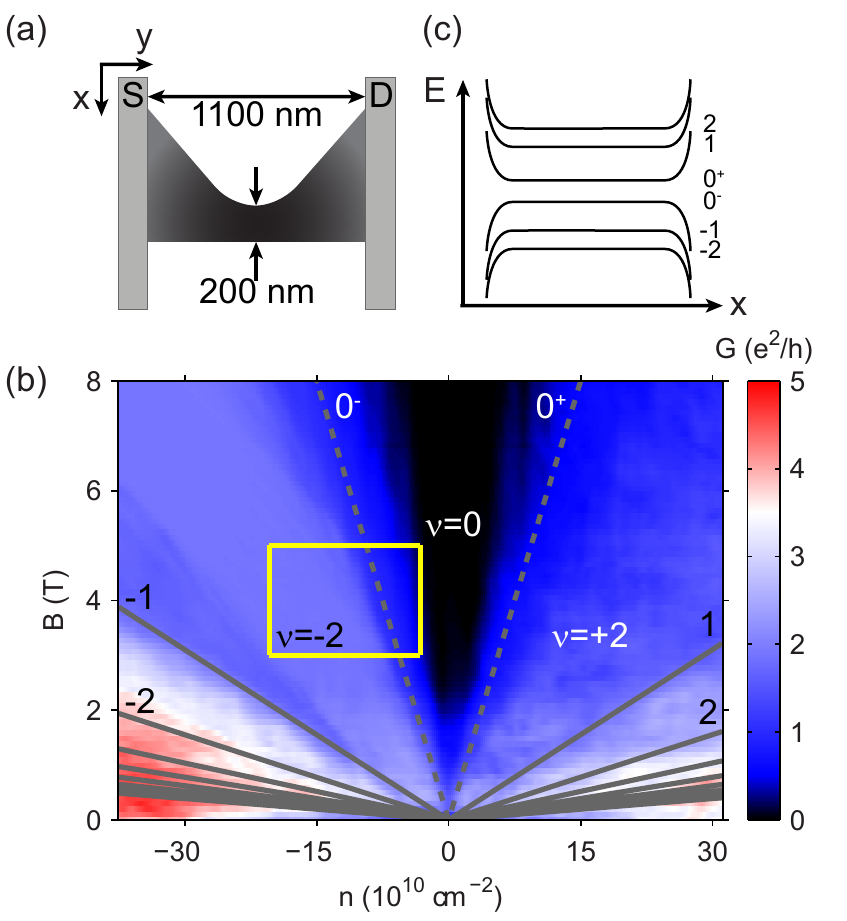}
\caption{(Color online) (a) An approximate geometry of the first measured nanoribbon, R1, and (b) its two-terminal conductance as a function of magnetic field and electron density. Grey lines mark the first few Landau levels (LLs). Dashed lines are the estimated positions of the twowise split 0th LL, indicated by $0^{-}$ and $0^{+}$. (c) Schematic of the $N=-2,-1,0^{-},0^{+},1,2$ Landau levels, including the split 0th level, with degeneracies of $g=4,4,2,2,4,4~[e B / h]$, respectively. 
}
\label{Figure1}
\end{figure}

{\it Measurements on a clean ribbon---}
We have fabricated suspended graphene nanoribbons, an approximate geometry of which is shown in Fig.~\ref{Figure1}a. We have used a polymer-based suspension method following Refs.~\citenum{Tombros2011, Maurand2014} and a transfer method by Ref.~\citenum{Dean2010}. Details are given in the Methods section. Measurements were done at 1.5~K using low frequency lock-in technique. 

Fig.~\ref{Figure1}b shows the two-terminal differential conductance $G$ of a nanoribbon (designated R1) as a function of the magnetic field $B$ and the electron density $n$, tuned by the gate voltage $V_{g}$. A conductance plateau takes shape at $\nu=-2$ filling factor, slightly below $G= | \nu | \rm e^{2}/h$ due to a contact resistance of $\rm 0.9 ~k \Omega$. Above 3~T, a widening zero-conductance region appears around the Dirac-point, caused by the splitting of the fourfold degenerate 0th Landau level due to finite-range Coulomb-interactions \cite{Zhang2006, Young2012, Yu2013, Abanin2013, Roy2014}. As confirmed by bias measurements, a true gap - in the order of 10~meV in this $B$-field range \cite{Roy2014} - forms between the upper and lower split 0th LL (denoted by indices $0^{+}$ and $0^{-}$), schematically shown in Fig.~\ref{Figure1}c.

\begin{figure*}[t!]
\centering
\includegraphics[width=1.0\textwidth,clip]{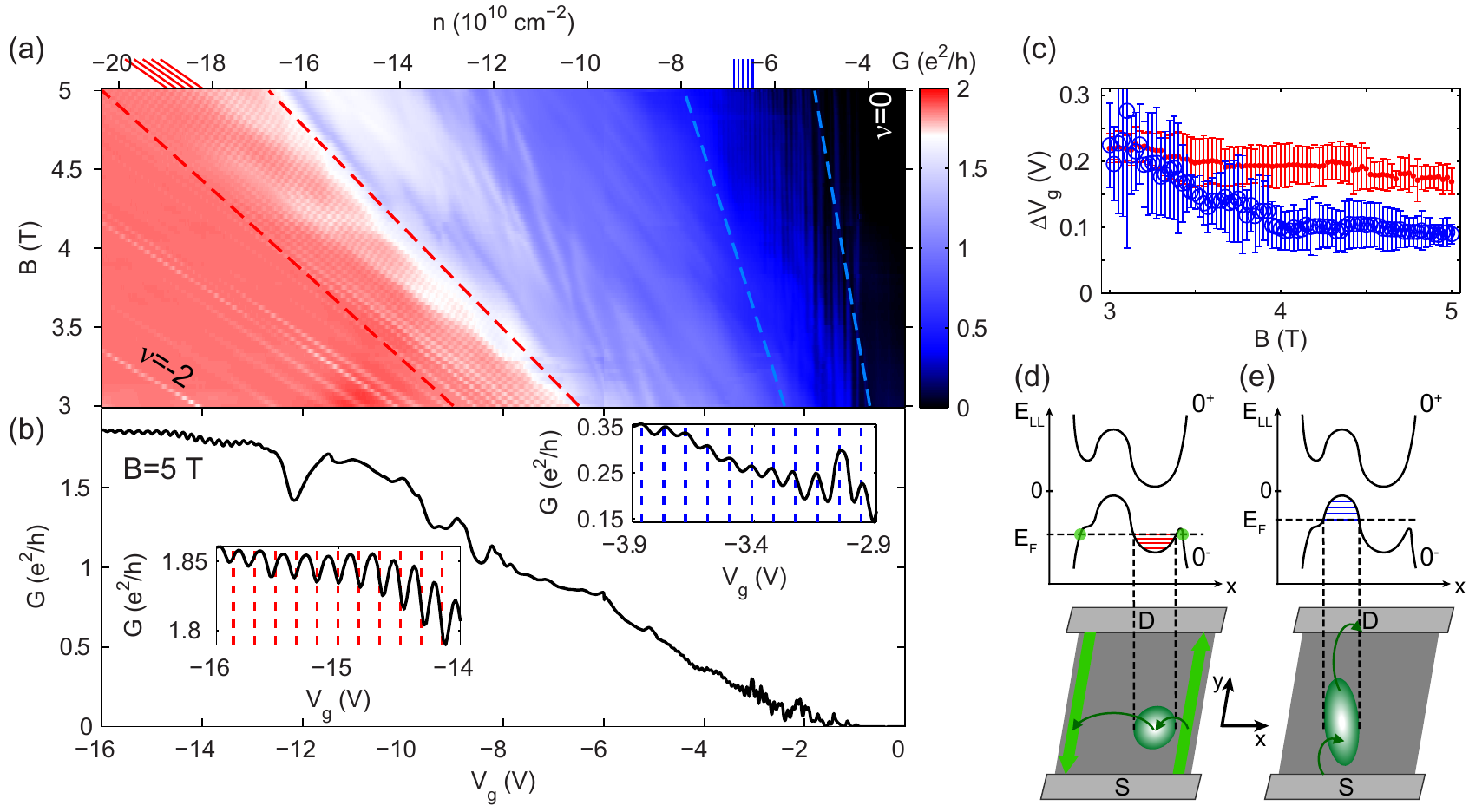}
\caption{(a) A zoom of the region highlighted by the yellow box in Fig.~\ref{Figure1}b. The observed fluctuation lines are either parallel with the center of the -2 plateau, or the center of the gap. The short red ($\nu=-2$) and blue ($\nu=0$) lines at the top indicate these directions of integer filling factors. Regular oscillations are found in the regions indicated by two pairs of dashed lines. (b) A slice of the map in (a) at $B=5$~T. Left and right insets are zooms of the regular plateau-edge and gap-edge fluctuations, respectively, which originate in the charging of a single quantum dot each. Dashed lines indicate periodicity. (c) Gate voltage period of the regular fluctuations averaged over the two regions of (a), i.e. between the red and blue pairs of dashed lines: dots (red) show the period of the plateau-edge, while empty circles (blue) show the period of the gap-edge oscillations.
(d), (e) The upper sketches depict the split 0th Landau level with a simplified disorder potential at a cross-section of the sample. Left side represents low filling, with an electron QD forming in a potential valley, while the right side shows high filling of the LL, with a hole QD defined by a potential hill. Horizontal red and blue lines represent the charging energy spacing of a dot. The left and right schematics in the lower half depict scattering between edge states or contacts/the rest of the dot network, causing conductance dips or peaks, respectively. The Coulomb-peak structure is especially recognizable in the left inset of (b), where it is inverted. 
}
\label{Figure2}
\end{figure*}

A zoom of the yellow rectangle in Fig.~\ref{Figure1}b can be seen in Fig.~\ref{Figure2}a, showing parts of the plateaus at $\nu=-2$ and $\nu=0$. The -2 plateau is separated from the gap by a wide transition region, where the $0^{-}$th LL is gradually filled, allowing scattering between edge states and contacts. A cut at 5~T (Fig.~\ref{Figure2}b) shows that the random conductance fluctutations visible in the transition region become very regular close to the gap or the $\nu=-2$ plateau. Zooms of these regions are shown in the insets of the same figure. These fluctuations are periodic in nature: at the plateau-edge, 18 oscillations can be seen with a period of 0.17~V, and at the gap-edge there are about 30 oscillations with 0.09~V spacing. We call the regular oscillations on the edge of the $\nu=-2$ plateau "plateau-edge oscillations", and the ones close to the $\nu=0$ region "gap-edge oscillations". Similar features were observed in the conductance band.

The plateau-edge and gap-edge oscillations are most visible between the red and blue dashed lines in Fig.~\ref{Figure2}a, and are parallel with the -2 and the 0 filling factor directions, respectively. These directions are marked with short red and blue lines at the top of the figure. Fig.~\ref{Figure2}c shows the oscillations' average periodicity as a function of magnetic field: red dots correspond to plateau-edge, while blue circles correspond to gap-edge oscillations. Their periodicity is approximately constant for a wide range of magnetic fields, except for the gap-edge oscillations' below 4~T, where the fluctuations become irregular. In the following, the mechanisms behind both random and regular conductance fluctuations, and their behavior, are addressed.

The transition region between the -2 and 0 plateaus points to a disorder potential that widens the $0^{-}$th LL in energy. In this region, the LL is partially filled, and the bulk is conducting due to delocalized states that connect edge states and contacts. Whereas near small, or almost complete filling, these states are localized to extrema of the potential landscape, stabilizing the quantum Hall plateaus. When a LL is almost empty, only the lowest disorder-potential valleys are filled with electrons, while in an almost full LL, the same happens in the hole picture. An example of the potential is shown in the top halves of Figs.~\ref{Figure2}d,~e. The conductance fluctuations observed near low and high filling may be resonant tunnelling events via the eigenenergy levels of the localized states. However, random potential features would produce eigenspectra that give random curves on the $n$-$B$ map \cite{Ilani2004, Martin2009}, contrary to the parallel, regular lines of the plateau and gap edges. 

The fluctuation lines' behavior is explained if we take electrostatic interactions into account. The disorder potential will be partially screened due to the electrons or holes present in the LL, which will accumulate in potential valleys and hills. Full screening, however, is not possible due to the limited number of states allowed within a given LL. The filled potential features result in a series of electron or hole islands with electrostatic charging energy, i.e. quantum dots, separated by tunnel barriers (Figs.~\ref{Figure2}d,~e), not unlike the finite $B$-field case of Ref.~\citenum{Allen2012}. Disorder-induced localized states have been visualized in 2DEGs and graphene using local probe techniques, such as single electron transistor \cite{Ilani2004, Martin2009}, scanning tunnelling microscopy and spectroscopy \cite{Miller2010, Jung2011} and spacially resolved photocurrent measurements \cite{Nazin2010}. Since these quantum dots cause scattering events between quantum Hall edge states and contacts (see the bottom halves of Figs.~\ref{Figure2}d,~e), their signature can be observed in conductance (Refs.~\citenum{Cobden1999, VelascoJr2010, Branchaud2010}, and even Refs.~\citenum{Zhang2009, McClure2009, Halperin2011}) and transconductance measurements \cite{Lee2012}.

The magnetic field dependence of the fluctuations, i.e. gathering together into sets of lines parallel with filling factor directions, is easily explained. Along a fluctuation line on the $n-B$ map, the average electron (hole) number on the originating dot is constant. Accordingly, the electron (hole) density belonging to the current LL is also constant, thus the fluctuations are parallel with the conductance plateau which corresponds to the empty (full) LL. 

For multiple dots, a random series of parallel lines is expected close to the plateau and gap edge, contrary to the periodic oscillations seen in the experiment in Fig.~\ref{Figure2}. Therefore, a single electron and hole QD must dominate scattering events for low and high filling of the $0^{-}$th LL, respectively. The questions arise, what makes a dot dominant, and in what circumstances? In the following, we give a physical picture and highlight the different mechanisms behind the plateau-edge and gap-edge oscillations. 

Since the plateau-edge oscillations have a negative contribution to the conductance plateau (see left inset of Fig.~\ref{Figure2}b), we infer that the dominant electron quantum dot connects mainly the edge states, causing backscattering. For a schematic drawing of the process, see Fig.~\ref{Figure2}d. In contrast, when we approach the gap, the $0^{-}$th LL is almost filled with electrons, and a single hole QD's charging dominates. In this case, however, no edge states exist, therefore the gap-edge oscillations can only result from forward scattering between the contacts (schematic in Fig.~\ref{Figure2}e). We attribute the dominant quantum dots to local potential extrema situated near the narrowest part of the ribbon, since conductance is most sensitive to this section. If the dominant dot were elsewhere, the observed oscillations would likely be irregular due to the contribution of other dots to either scattering mechanism. Nonetheless, for the hole QD, the rest of the dot network - in the wider sections of the sample - is essential to establish a connection toward the contacts.

The estimation of the electron and hole dots' sizes supports this suggestion. The exact capacitance per area can be calculated from the slope of the -2 plateau center, since the filling factor relates to the density via $\nu=hn/eB$. We calculate the capacitance per area, $\widetilde{C}= dn / dV_{g}$, to be $\rm ( 1.07 \pm 0.02 ) \cdot 10^{10}~\rm cm^{-2}/V$, which agrees well with electrostatic calculations on similar devices \cite{Rickhaus2015, Liu2015}. Using the average gate voltage periodicities $\Delta V_{g}$ in Fig.~\ref{Figure2}c, we estimate that the electron quantum dot responsible for the plateau-edge oscillations extends over an area of approximately $( \widetilde{C} \cdot \Delta V_{g} )^{-1} = ( 4.7 \pm 0.5 ) \cdot 10^4$~nm$^{2}$, while the hole quantum dot - causing the gap-edge oscillations - is $( 9.3 \pm 1.9 ) \cdot 10^4$~nm$^{2}$ in size. Since the ribbon's narrowest part is $\sim200$~nm wide, quantum dots with the above areas are able to cause scattering events across the width or length of the constriction, thus connecting the edge states or the wider, highly doped regions (and therefore the contacts). 

Since the dominant QDs are formed in potential extrema of the constriction, their signatures are best seen at either low or high Landau level filling, between the dashed lines in Fig.~\ref{Figure2}a. Moving the Fermi level toward the LL's center, more potential valleys or hills start to play a role in transport, and the fluctuation pattern becomes random. Eventually, all charges become delocalized, and the bulk becomes conducting. In contrast, increasing the magnetic field while following a fluctuation line from its high-visibility region between the red (blue) dashed lines of Fig.~\ref{Figure2}a, and into the -2 (0) plateau, the size and coupling of localized and edge states decreases. Tunnelling rates are suppressed, and eventually only the flat plateau remains visible.

\begin{figure} [b!]
\centering
\includegraphics[width=0.45\textwidth,clip]{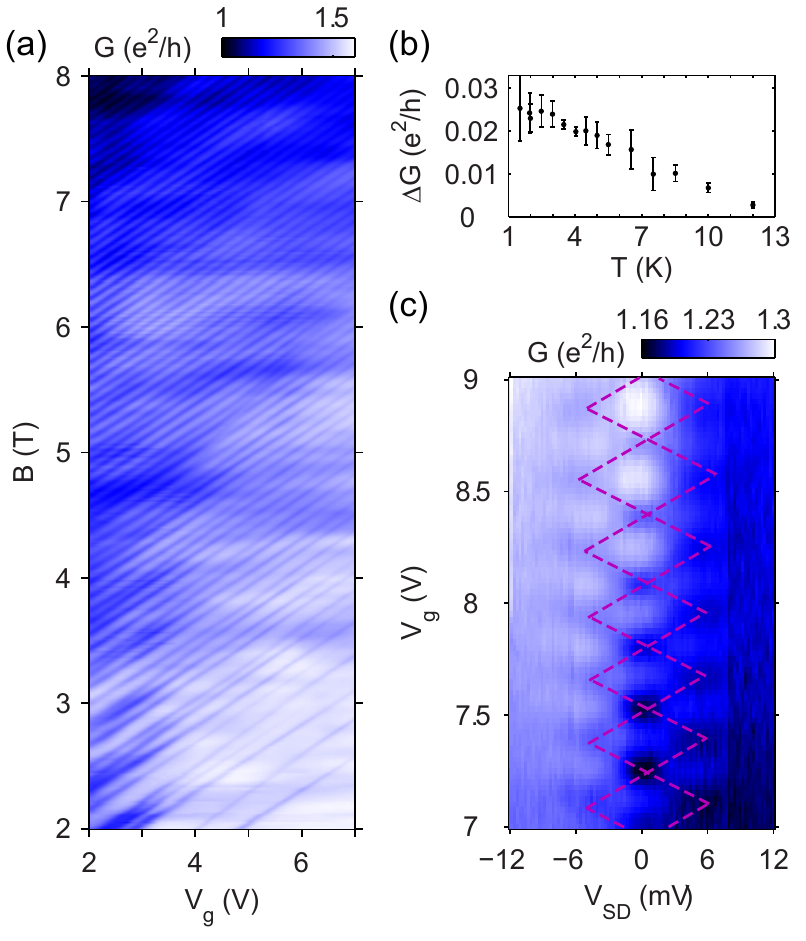}
\caption{(a) Conductance of nanoribbon R2 in the electron regime. (b) The average peak-to-peak amplitude of the periodic oscillations between $V_g=7$ and 9~V, at 8~T, as a function of temperature. (c) Conductance at 8~T as a function of DC bias and gate voltage. Dashed lines are guides to the eye for the Coulomb-diamond structure.
}
\label{Figure3}
\end{figure}

{\it Coulomb peak behavior---}
We reproduced the oscillation pattern in the conductance band of a second nanoribbon, designated R2, that didn't show well-developed quantum Hall plateaus. Its two-terminal conductance, displayed in Fig.~\ref{Figure3}a, shows regularly placed conductance dips, their lines parallel with the expected slope of the $\nu=+2$ plateau. Therefore, they are attributed to a hole dot belonging to the electron side of the 0th LL, causing backscattering. Fig.~\ref{Figure3}b shows their peak-to-peak amplitude at 8~T as a function of temperature. The fluctuations disappear in the $10-15~K$ range, where the charging energy of the QD becomes comparable with thermal broadening. Since the oscillations' width is similar to their period, fitting curves on the series of inverted Coulomb-peaks, or their amplitude - to analyze height and width change with temperature - can't be done without a huge margin of error.

The fluctuations' slope, parallel with the $\nu=+2$ direction, gives the gate capacitance per area, which is $\widetilde{C} =( 1.36 \pm 0.04 ) \cdot 10^{10}~\rm cm^{-2}/V$, or $21.8 \pm 0.6~ \rm aF/ \mu m^{2}$. With the oscillation period, we estimate the area of the dot to be approximately $( 2.7 \pm 0.3 ) \cdot 10^4$~nm$^{2}$. In Fig.~\ref{Figure3}c the dominant quantum dot's stability diagram, i.e. conductance versus $V_{g}$ and $V_{SD}$ (source-drain voltage), is shown. The conductance contribution of the Coulomb-diamonds is negative, since the dot causes backscattering. Their size gives a charging energy of $5.5 \pm 0.5 ~ \rm meV$, allowing us to calculate the self capacitance: $C_{\Sigma}=29.1 \pm 2.6 \rm ~aF$. As a comparison, the gate capacitance is $C_{g}=0.58 \pm 0.07 \rm ~aF$. By counting the number of regular oscillations, we estimate that the height of the potential hill that defines the hole QD is a remarkable 260~meV, comparable to the energy of the first LL at 8~T, $\sim100$~meV. However, the charging energy deduced from the size of the Coulomb diamonds in the source-drain axes might be overestimated, since not all of the bias voltage drops at the barriers defining the quantum dot.

\begin{figure} [t!]
\centering
\includegraphics[width=0.5\textwidth,clip]{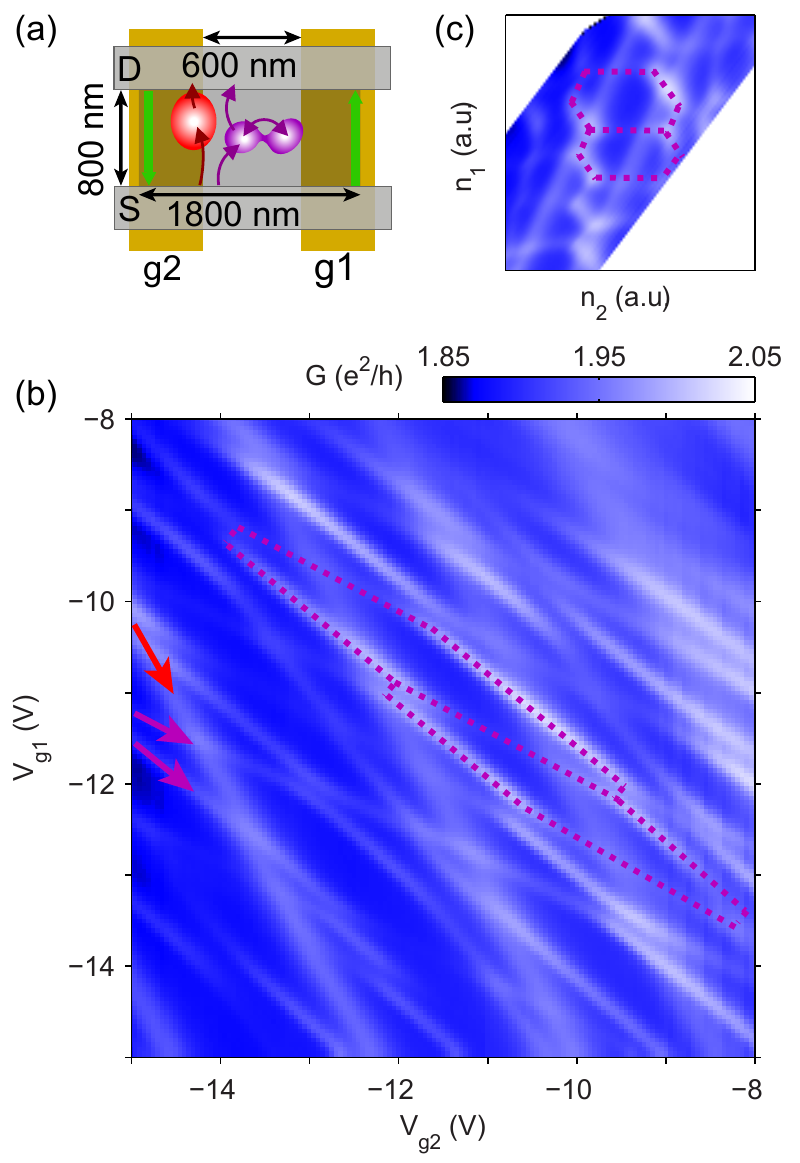}
\caption{ (a) A graphene strip tuned by two bottom gates, whose conductance at 4~T is displayed on (b), with both sides of the sample tuned near the edge of the $\nu=-2$ plateau. A two-dot subsystem's charge stability diagram can be recognized on a background that originates from other dots. The hexagons of purple dashed lines are guides to the eye. A schematic of such a QD arrangement can be seen in (a). (c) The same as (b) but compensated for cross-capacitances.
}
\label{Figure4}
\end{figure}

{\it Double-dot system in a wide sample---}
To examine the role of sample width, we measured the conductance of a 1.8~$\mu$m wide and 0.8~$\mu$m long graphene strip. The density of the device could be locally tuned by two bottom gates, g1 and g2, that were aligned in parallel with the sample current direction, as shown in the schematic in Fig.~\ref{Figure4}a.

Fig.~\ref{Figure4}b shows the conductance of the quantum Hall plateau near $\nu=-2$ filling of both sides, as a function of the two gate voltages, at 4~T. A random structure of lines with different slopes - some of them highlighted by arrows - are conspicuous on the conductance map, indicating that QDs are tuned by both voltages. The slope of a dot's fluctuation lines is determined by the dot's position relative to the two gate electrodes. As expected, the map shows the signatures of a network of QDs. Due to the low aspect ratio, scattering between contacts is much more likely than between edge states, explaining the positive conductance contribution of the dots. 

Some of the fluctuation lines show avoided crossings. They have similar slopes, indicating they belong to quantum dots that are close to each other, enabling them to hybridize (purple QDs in Fig.~\ref{Figure4}a). Thus, the lines with avoided crossings belong to one or more double-dot systems. The expected hexagonal pattern of a double dot is highlighted in purple as a guide to the eye, and is even more evident in Fig.~\ref{Figure4}c, where the map is distorted to compensate for cross-capacitances. This way the conductance is shown as a function of the individual dot charges of this double-dot system. One set of lines is stronger, suggesting one dot is better coupled to the contact electrodes than the other.

{\it Summary and outlook---}
A band gap is essential to create graphene transistors and spin qubits. However, Klein tunnelling limits the effectiveness of electrostatic confinement, while hard wall confinement (etching) introduces further obstacles. In the quantum Hall regime, a disorder potential can act as confinement due to the bulk gaps between Landau levels. As a result, a network of quantum dots form. In our nanoribbons the small sample width enabled a single QD to dominate that could be read out not only by contacts, but also by edge channels. In a wide flake with two gates a double-dot system's hexagonal pattern was observed. This mechanism suggests a way to electrostatically confine electrons in clean single-layer graphene devices using multiple gate electrodes. With suitable geometries, the creation of quantized conducting channels, single and double quantum dots, quantum point contacts, and even interferometers becomes possible.

{\it Acknowledgements---}
We acknowledge useful discussions with Romain Maurand, Andreas Baumgartner, Andr\'{a}s P\'{a}lyi, P\'{e}ter Rakyta, L\'{a}szl\'{o} Oroszl\'{a}ny, Matthias Droth, and Csaba T\H{o}ke. This work was funded by the EU ERC CooPairEnt 258789, Hungarian Grants No. OTKA K112918, and also the Swiss National Science Foundation, the Swiss Nanoscience Institute, the Swiss NCCR QSIT, the ERC Advanced Investigator Grant QUEST and the EU Graphene Flagship project. S.C. was supported by the Bolyai Scholarship.

{\it Methods---}
Fabrication steps followed Refs.~\citenum{Tombros2011, Maurand2014}. First, 5/55~nm thick Pd/Al or Ti/Au bottom gates were fabricated on a p:Si/SiO$_{2}$ layer, which were covered first with a 50~nm ALD-grown $\rm Al_{2}O_{3}$ insulating layer, second with 600~nm thick LOR resist. Graphene was exfoliated onto a separate wafer and transferred using the method described in Ref.~\citenum{Dean2010}. Subsequently, the flake was contacted with 40~nm thick Pd wires, and etched using e-beam lithography and reactive ion etching. Approximate dimensions of ribbons R1, R2 are given in Fig.~\ref{Figure1}a. Finally, graphene was suspended by exposing and developing the LOR resist below. Samples were current annealed at low temperature to remove solvent and polymer residues. Measurements were carried out at 1.5~K, using standard lock-in technique. The Dirac-points of the ribbons R1, R2, and the wide sample were at approximately $V_{G} \approx 3$,~0, and 1~V, respectively.





\end{document}